\documentclass[conference]{IEEEtran}
\usepackage{cite}
\usepackage{graphicx}
\usepackage{amsmath}
\usepackage{bbm}
\usepackage{hyperref}
\usepackage{url}
\usepackage{tikz}

\usepackage{bm}
\renewcommand{\vec}[1]{\bm{#1}}
\newcommand{\indicator}[1]{\mathbbm{1}_{\{#1\}}}

\IEEEoverridecommandlockouts

\newcommand\copyrighttext{%
\footnotesize \textcopyright \enspace 2018 IEEE. Personal use of this material is permitted. Permission from IEEE must be obtained for all other uses, in any current or future media, including reprinting/republishing this material for advertising or promotional purposes, creating new collective works, for resale or redistribution to servers or lists, or reuse of any copyrighted component of this work in other works.  DOI: \href{https://doi.org/10.1109/BlackSeaCom.2018.8433721}{10.1109/BlackSeaCom.2018.8433721}
}
\newcommand\copyrightnotice{%
\begin{tikzpicture}[remember picture,overlay]
\node[anchor=south] at (current page.south) {\fbox{\parbox{\dimexpr\textwidth-\fboxsep-\fboxrule\relax}{\copyrighttext}}};
\end{tikzpicture}%
}

\begin{document}

\title{Joint Power Control and Time Division to Improve Spectral Efficiency in Dense Wi-Fi Networks \thanks{The research was done at IITP RAS and supported by the Russian Science Foundation (agreement No 16-19-10687).}}

\author{\IEEEauthorblockN{
	Evgeny Khorov\IEEEauthorrefmark{1}\IEEEauthorrefmark{2},
	Anton Kiryanov\IEEEauthorrefmark{1},
	Alexander Krotov\IEEEauthorrefmark{1}\IEEEauthorrefmark{2}
}
\IEEEauthorblockA{ \IEEEauthorrefmark{1}Institute for Information Transmission Problems, Russian Academy of Sciences, Moscow, Russia\\
\IEEEauthorrefmark{2}Moscow Institute of Physics and Technology, Moscow, Russia \\
Email: \{khorov, kiryanov, krotov\}@iitp.ru}
}

\maketitle
\copyrightnotice
\begin{abstract}
Ubiquitous densification of wireless networks has brought up the issue of inter- and intra-cell  interference. Interference significantly degrades network throughput and leads to unfair channel resource usage, especially in Wi-Fi networks, where even a low interfering signal from a hidden station may cause collisions or block channel access as it is based on carrier sensing. In the paper, we propose a joint power control and channel time scheduling algorithm for such networks, which significantly increases overall network throughput while maintaining fairness. The algorithm is based on branch-and-bound global optimization technique and guarantees that the solution is optimal with user-defined accuracy.
\end{abstract}

\section{Introduction}

For the last years, the density of wireless networks has grown significantly, which leads to huge interference both between overlapped networks and between stations (STAs) within a single network. Constant desire to increase transmission rates and to provide higher Quality of Experience (QoE) to end users can no longer be fulfilled only by means of designing faster Modulation and Coding Schemes (MCSs) and increasing the number of antennas at transmitter and receiver. Today, interference is a critical issue which reduces the performance of wireless networks. To cope with this problem in Wi-Fi networks,  IEEE~802 LAN/MAN Standards Committee has formed a new task group which aims to design a new amendment to the Wi-Fi standard, namely IEEE~802.11ax. As listed in the corresponding Project Authorization Request document~\cite{ieee802.11axTG}, a goal of the new amendment is at least 4x~increase in throughput in dense deployment scenarios. While many other features of IEEE~802.11ax --- such as OFDMA or uplink multi-user MIMO --- have been quickly adopted, interference mitigating solutions brought up many questions and only few of dozens proposal (the most general ones) got enough votes to become standardized. A lot of proposals aimed at enabling spatial reuse and increasing throughput in dense networks were presented at IEEE 802.11 meetings but most of them were rejected. The prevalent approaches include dynamic control of transmission power and carrier sense threshold (CST). Apart from that Time Division Multiple Access (TDMA) is also considered as a mean to improve performance in scenarios with hidden STAs. Fortunately, in IEEE~802.11ax the access point (AP) can schedule channel time and transmission parameters for both uplink and downlink traffic.

In this paper, we combine the aforementioned approaches, state a global optimization problem and design a hybrid algorithm to solve it. Specifically, in Section~\ref{sec:previous}, we review prior art. Section~\ref{sec:problem} describes our scenario of interest and states mathematically global optimization problem. In Section~\ref{sec:algorithm}, we design a novel algorithm which solves the stated problem. In Section~\ref{sec:results}, we describe numerical results. Section~\ref{sec:conclusion} concludes the paper.

\section{Related Papers}
\label{sec:previous}

A number of algorithms tuning transmission parameter values were proposed in literature in attempt to decrease interference in dense wireless networks.

Several existing approaches tune CST.
In \cite{zhu2004adapting}, the authors demonstrate that properly tuned CST can improve spatial reuse while avoiding hidden and exposed node problems without using virtual carrier sensing.
Significant disadvantage of their algorithm is that data rates and transmission powers are assumed to be selected in advance. Authors of \cite{fuemmeler2006selecting} propose to tune both transmission power and CST. The main idea is to keep the product of transmission power and CST constant to avoid unfairness.
To simplify the problem, many existing papers either (i)~assume that the noise is negligible~\cite{zhu2004adapting}, or (ii)~do not take into account restrictions on maximum power and CST~\cite{fuemmeler2006selecting}.
From mathematical point of view both assumptions are equivalent. If there is no noise, all transmission powers can be proportionally reduced to meet any requirements on maximum transmission power and CST. Vice versa, if there are no restrictions on maximum transmission power and CST, it is possible to increase transmission power at all the STAs until noise becomes negligible.
However, in real devices transmission power is limited by physical and regulatory constraints. Apart from that, in unlicensed spectrum, the STAs have either to transmit very rarely (with a quite long duty cycle) or to follow listen before talk principle and a restriction on CST. For these reasons, we need an algorithm which takes into account a restriction on CST in a scenario with nonzero noise power.

In \cite{zhou2007joint}, the authors propose a joint CST, transmission power and MCS tuning algorithm based on several assumptions.
First, there are no hidden STAs before tuning. Second, noise power is \emph{much higher} than interference from neighbor STAs.
Third, there is a known number of interfering transmissions \(k\)
(either 4 or 6)
that contribute to interference almost equally, i.e. their method only works when the network topology is regular. Transmission power is selected as the minimum value that can guarantee a particular data rate given \(k\) interfering STAs. As well as~\cite{fuemmeler2006selecting}, the authors keep constant the product of transmission power and CST threshold. The developed algorithm is decentralized and executed independently for each transmitter-receiver pair. It estimates SNR
at the receiver and selects the highest data rate that can be supported in the absence of interference
Then it tunes transmission power to achieve SINR required by the selected rate. Decentralization of the algorithm is an advantage, but optimal solution is not guaranteed due to multiple
assumptions. Therefore, we turn to the centralized power control problem.

Some classic works on centralized power control are concerned with achieving specific SINR values~\cite{stefanyuk1967collective,foschini1993simple}.
At those times, communication networks were mostly used to transmit voice and it was not necessary to increase SINR as high as possible. In~\cite{stefanyuk1967collective} the problem of choosing transmission powers given exact SINR requirements is considered. Authors show that the problem statement can be extended to wireless networks with relays. To find transmission powers, they construct a system of linear equations, solving which they obtain a transmission power vector corresponding to the given SINR vector. They consider a set of vectors for which this linear system could be solved. The authors show that this set is connected.
For that, they also prove that for each given point in this set it is possible to decrease SINR for any particular receiver without changing SINR on any other receiver. The proposed distributed power control algorithm consists in reducing power for pairs of STAs which have SINR higher than the required and increasing power for STAs with SINR lower than the required. The algorithm is proven to be stable given that problem is feasible. The authors also propose an extension to this algorithm. STAs try to reach the required SINR or utility which is a monotonous function of SINR (e.g., error probability) and completely disable themselves if  they do not achieve the required SINR or utility value at the maximum transmission power. Thus, the requirements can be satisfied for a subset of users if the problem is not feasible for all users.
Described above problem is a linear optimization problem and can be easily solved even by distributed algorithms. Authors of~\cite{foschini1993simple} develop a distributed algorithm which achieves target SINR for all users provided that it is feasible.

For elastic traffic (such as web browsing and adaptive streaming), it is important to maximize data rate. Data rate maximization is a non-convex optimization problem, so even the development of a centralized algorithm is a much more complicated task.
The problem of maximizing network throughput by power control is considered in \cite{chiang2007power}. The authors show that power control is a nonlinear optimization problem. To simplify it, they adopt the assumption that SINR at receivers is much greater than 1. Under such an assumption a non-convex optimization problem can be reduced to a convex optimization problem. The drawback of this approach is that the assumption of high SINR does not hold in dense networks. To eliminate this drawback, the authors propose a heuristic that solves the problem when SINR is $\approx 0$ dB, which, however, does not guarantee optimality of the found solution.

Studying the problem of power allocation for a set of STAs, the authors of \cite{qian2009mapel} develop an algorithm called MAPEL (Multiplicative Linear Fractional Programming-bAsed PowEr ALlocation), which can find the global optimal solution of the weighted throughput maximization.
MAPEL is based on the polyblock algorithm, which is an algorithm for solving monotonic optimization problems\cite{tuy2016convex}.
The main drawback of MAPEL (inherent in the polyblock algorithms) is extremely slow convergence if in the optimal case, one of the transmission powers is close to zero. This happens because with the CST constraints, it is sometimes necessary to completely disable a transmitter for most of the time to achieve global optimization.
Another drawback of MAPEL is that it does not consider the ability to avoid interference by scheduling interfering STAs in different channel times. This issue is partially addressed in~\cite{qian2010smapel}, which solves joint power control and scheduling problem. A new algorithm called \mbox{S-MAPEL} reduces dimension of the problem to finite number of variables. However, even with finite dimension, the algorithm complexity is too high, so the authors propose an accelerated version of the algorithm, \mbox{A-S-MAPEL}, based on the idea that points with equal utilities are likely to be symmetric, i.e. differ only in the order of schedule. They compare their solution to on-off scheduling, when power control is simplified and STAs can only transmit at maximum power or not transmit at all. In this case, it is possible to achieve results which are not significantly worse than A-S-MAPEL results. However, the complexity of this problem can be bypassed by solving scheduling problem dynamically, as we show in Section~\ref{sec:algorithm}.

Slow convergence problem can be avoided with the branch-and-bound approach, as used in \cite{weeraddana2011weighted}.
Authors of  \cite{weeraddana2011weighted} apply this approach to solve power control problem in the same scenario as in \cite{qian2009mapel} without CST constrains.

In this paper, we introduce a new algorithm that extends the branch-and-bound approach to take into account CST requirements and combines power control with channel time scheduling to achieve fairness between users.

\section{Problem Statement}
\label{sec:problem}
\subsection{Scenario}
	
Consider a network consisting of \(N\) pairs of transmitters and receivers. Each traffic flow is established between pair $i$ of nodes, i.e. transmitter $i$ and receiver $i$. Transmitter $i$ emits power $x_i$. Receiver \(i\) has a thermal noise level \(n_i\). We assume that data transmissions are significantly longer than acknowledgment transmissions and consider only unidirectional data transmissions. Such an assumption is especially natural for block acknowledgments --- introduced in IEEE~802.11n --- when multiple data frames are acknowledged with a single control frame. Besides, acknowledgments are transmitted using robust MCS and thus they are usually not damaged.

Network topology is described by pathloss attenuation between transmitters and all other nodes.
Let $a_{ij} \geq 0$ be a pathloss attenuation between transmitter \(j\) and receiver \(i\),
and $b_{ij} \geq 0$ be a pathloss between transmitter \(j\) and transmitter \(i\).
Note that all receivers are able to receive signal from corresponding transmitter, so
$a_{ii} > 0, \forall i.$ Transmitters do not interfere with themselves: $b_{ii} = 0, \forall i$.
In general case, the network may contain repeaters,  which can be modeled as pathloss attenuation greater than \(1\) (i.e. pathloss ``gain'')~\cite{stefanyuk1967collective}.

\subsection{Power Control and Scheduling Problem}
\label{sec:powercontrol}

The aim of a power control algorithm is to maximize a system utility function $\hat U$ by choosing transmission power \(x_i\) for each transmitter \(i\). User utility function $U$ is usually a monotonous function of its rate $r_i$, in particular, in this paper we consider \(\alpha\)-fairness~\cite{Mo2000FairEW}:
\begin{equation}
\label{eq:utilit}
U(r_i) = \begin{cases}
\log(r_i), & \alpha = 1, \\
\frac{r_i^{1-\alpha}}{1 - \alpha}, & \alpha \ge 0, \alpha \ne 1.
\end{cases}
\end{equation}

\noindent $\hat U(\vec r)$ is a weighted sum of user utility functions:
\begin{equation}
\hat U(\vec r)= \sum_i w_i U(r_i),
\label{eq:power_utility}
\end{equation}
where $\sum_i w_i = 1$.

Data rate is a non-decreasing function of SINR $\gamma_i$: $r_i = f(\gamma_i(\vec x)),$ which can be estimated according to error rate models of available modulation and coding schemes, while SINR at receiver \(i\) is calculated as follows:

\begin{equation}
\label{eq:gamma}
	\gamma_i(\vec x)= \frac{a_{ii} x_i}{n_i + \sum\limits_{j \neq i} a_{ij} x_j}.
\end{equation}

The described optimization problem is subject to several restrictions.
First, transmission power \(x_i\) cannot exceed \(\hat x_i\). Second, each STA $i$ operates in unlicensed spectrum and following listen-before-talk rule cannot transmit if its received power exceeds CST \(c_i\), while \(c_i\) must not exceed regulatory threshold \(\hat c\). Since lowering \(c_i\) can only limit the set of feasible solutions, hereinafter we suppose $c_i = \hat c, \forall i$.
Noise can be accounted for by reducing CST, but thermal noise is typically negligible compared to CST and can be simply ignored.
Thus transmission is allowed\footnote{For simplicity, we suppose that this condition shall hold during the whole transmission, though in real networks it is checked only before starting a new transmission. Thus, we forbid to start a new transmission if it violates the above condition for any \emph{already started} transmission.} only if
$	\sum_j b_{ij} x_j  \le  c_i, \forall i.$

The formal problem statement is as follows:
\begin{equation*}
	\begin{aligned}
		& \underset{\vec{x}}{\text{maximize}}
		& & \sum_i w_i U\left(f\left(\frac{a_{ii} x_i}{n_i + \sum\limits_{j \neq i} a_{ij} x_j}\right)\right),
	\end{aligned}
\end{equation*}
\begin{equation}
\label{eq:cst}
	\begin{aligned}
		& \text{subject to}
		& & \sum_j b_{ij} x_j \le \hat c, \forall i \text{ such that } x_i>0; \\
		& & & 0 \le x_i \le \hat x_i, \forall i.
	\end{aligned}
\end{equation}

If two pairs of transmitting and receiving nodes are located close to each other, it is worth to alternate transmissions in different pairs. While mentioned in Section \ref{sec:previous} papers on joint power control and scheduling consider only static schedule that has high computational complexity and does not take into account possible fluctuations of traffic intensity and channel conditions, in this paper we allocate powers every time unit, taking into account the results of previous decisions.

\section{Proposed Algorithm}
\label{sec:algorithm}

\subsection{General Idea}

Existing polyblock-based algorithms, such as MAPEL,
are known to converge slowly when some variables of solution are close to zero\cite{tuy2016convex}.
With the CST restriction, optimal data rates for some transmitters can be zero.
For this reason, instead of polyblock, we use the branch-and-bound approach.

The idea of the branch-and-bound algorithm is to cover the whole feasible set with a \emph{box}, which is a Cartesian product of  intervals covering feasible values of each variable, and subsequently divide (branch) it into smaller boxes and remove boxes which either have utility function value lower than already achieved or are infeasible.

We operate with vectors of rates $\vec r$ as variables instead of vectors of powers $\vec x$, because the  branch-and-bound approach relies on the fact that the utility function is a monotonous function of the variables being optimized.
A box is determined by two vectors  $\vec p$ and $\vec q$ corresponding to the minimal and maximal rates $\vec r$ within the box. Thus, we represent the box as \([\vec p, \vec q]\). Given $\vec r$, we can obtain the vector of required SINRs $\vec{\gamma}$ using $f^{-1}(\cdot)$. Then we can find $\vec x$ by solving the system of linear equations \eqref{eq:gamma} subject to \eqref{eq:cst} or conclude that no solution exists and such $\vec r$ is infeasible.

The algorithm maintains a set of boxes. To each box, we assign estimated upper bound $U_{max}$, which can be lower than $\hat U (\vec q)$.
Apart from that, the algorithm maintains the best found feasible solution and the corresponding utility value $U_{best}$.

Branch and bound is an iterative algorithm. At each iteration, the algorithm selects a box with the highest $U_{max}$ and performs branching (Section~\ref{sec:branch}), reduction (Section~\ref{sec:reduction}) and bounding (Section~\ref{sec:bounding}). The algorithm terminates when it finds a solution close enough to the optimal one in terms of average data rates.

\subsection{Initialization}
\label{sec:init}
First of all, we consider the utopia point  corresponding to the vector of rates achieved if there were no interference:
\begin{equation}
f\left(\frac{a_{ii} \hat x_i}{n_i}\right),
\label{eq:utopiatates}
\end{equation}

Since $f$ is a limited function, the utopia point may be feasible. In such a case, we have found the desired solution and the algorithm stops.
Otherwise, we create a box that covers the set from zero to the utopia point.

For the initial box, the best found solution is initialized with zero rates \(\vec r\). The upper bound  $U_{max}$ is set to the utility value of the utopia point.

\subsection{Branching}
\label{sec:branch}

The selected box is split into two boxes along the longest side. Both the resulting boxes are reduced as described in the next section, bounded and placed back into the box set.

\subsection{Reduction}
\label{sec:reduction}

During the reduction step, we refine the limits of each variable within the box by removing infeasible points or points with the utility lower than $U_{best}$. If the box definitely does not contain points with the utility higher than $U_{best}$, the box is removed from further consideration.

Consider box $[\vec p, \vec  q]$.
If $U(\vec p)$ is less than $U_{best}$, we try to move the lower boundary of the box to exclude points which cannot provide utility higher than already obtained one. To determine the new lower boundary of the box, we repeat the following procedure for each component $p_k$. Firstly, we build vector $\vec r$ of rates, for which $r_i=q_i$, $\forall i \ne k$, i.e. it is the highest possible rate in this box. Secondly, we find the lowest $r_k$ for which the system utility is not lower than \(U_{best}\) by solving equation

\begin{equation}
	w_k U(r_k) + \sum_{i \neq k} w_i U_i(q_i) = U_{best}.
	\label{eq:ak}
\end{equation}
Obviously, all vectors from the considered box $[\vec p, \vec q]$ with the $k$-th component less than $r_k$ cannot give utility higher than $U_{best}$. So if $r_k>p_k$, we replace $p_k$ with $r_k$.

Similarly, we reduce the upper boundary by solving
\begin{equation}
w_k U(r_k) + \sum_{i \neq k} w_i U_i(p_i) = U_{max}.
\label{eq:bk}
\end{equation}
Note that $U_{max}$ can be lower than $U(\vec q)$ because of the bounding step described below.

\begin{figure*}[!ht]
	\centering
	\includegraphics[width=7in]{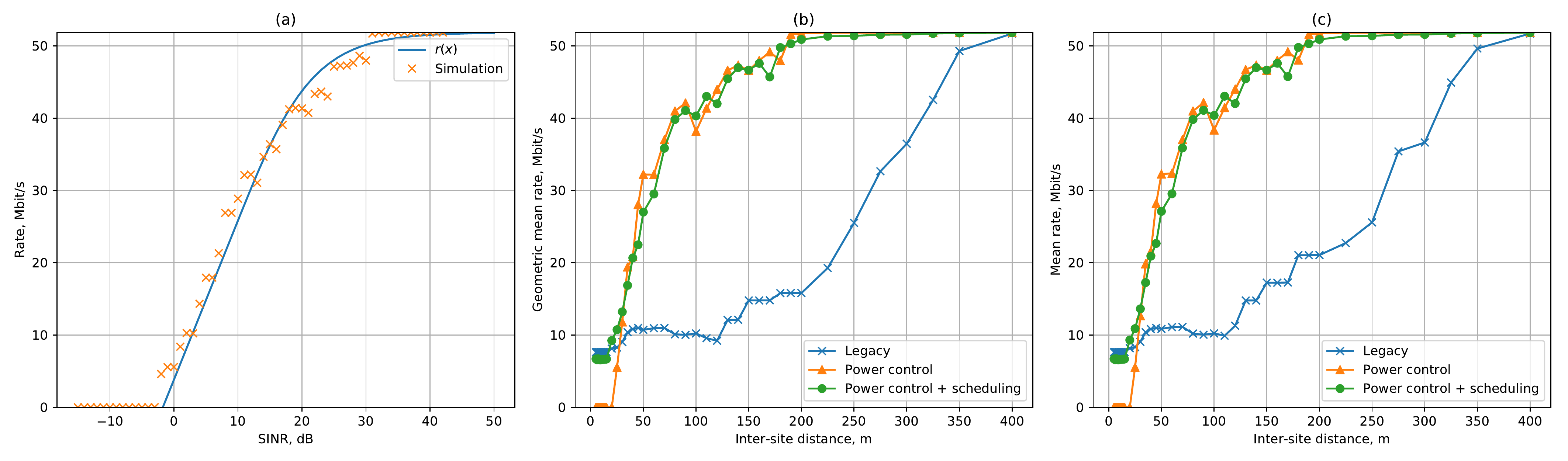}
	\caption{(a) Data rate as a function of SINR; (b) Geometric mean and (c) Mean data rate as  a function of inter-site distance}
	\label{fig:rate}
	\label{fig:utility}
	\label{fig:throughput}
\end{figure*}

\begin{figure}[!ht]
	\centering
	\includegraphics{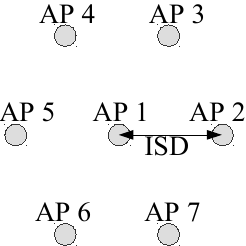}
	\caption{Network topology}
	\label{fig:topology}
\end{figure}

\subsection{Bounding}
\label{sec:bounding}

The goal of bounding step is to reduce upper bound $U_{max}$ for the considered box \([\vec p, \vec q]\) using bisection method if $\vec q$ is not feasible. Otherwise, $U_{max}=\hat U(\vec q)$.

Specifically, we introduce auxiliary variables  \(\vec l\) and \(\vec h\). Initially, let  \(\vec l = \vec p\), \(\vec h = \vec q\). We repeatedly find the middle point \(\vec m = \frac{1}{2} \vec l + \frac{1}{2} \vec h\) and check its feasibility by converting rates to the corresponding  values of SINR and solving the system of linear equations \eqref{eq:gamma} subject to \eqref{eq:cst}.
If $\vec m$ is feasible, it replaces \(\vec l\), otherwise it replaces \(\vec h\). We continue such a procedure until \(U^{-1} (\hat U(\vec h)) - U^{-1} (\hat U(\vec l))\) is below \(\varepsilon\), which is the target accuracy expressed in Mbit/s.

To update the upper bound $U_{max}$ in the feasible part of box $[\vec p, \vec q]$, we should exclude box $[\vec h, \vec q]$, since it is infeasible. Then we consider $N$ points with the same components as \(\vec q\) except one component, which is replaced with the corresponding component of \(\vec h\).
Since $U$ is a monotonic function of rate, it is easy to show that
at least one of the aforementioned  points is the maximum point of $U$ in the remaining part of box $[\vec p, \vec q]\setminus [\vec h, \vec q] $.
Let $U^*$ be the maximal value of the utility function in these points. If $U^*$ is less than the current value $U_{max} $, we replace $U_{max} $ with $U^*$.

Apart from that, if \( U(\vec l) > U_{best}\), we set \(U_{best} = U(\vec l)\) and update the best found solution accordingly, since \(\vec l\) is feasible.

\subsection{Termination}

Each time the algorithm selects a box for the branching step, it terminates if no solution from this box can significantly improve the best found solution, i.e. if:
\begin{equation}
\label{eq:term}
	 U^{-1}(U_{max}) < U^{-1}(U_{best})+ \varepsilon.
\end{equation}
Note that since the box with the highest $U_{max}$ is always selected, all the remaining boxes will also satisfy \eqref{eq:term}.

\subsection{Dynamic Scheduling}
\label{sec:dynamic}

The idea of dynamic scheduling is to periodically run power control algorithm described in Section~\ref{sec:powercontrol}, taking into account the rate history. For that, we consider alpha-fairness utility function~\eqref{eq:power_utility} with weights selected as follows~\cite{schwarz2011throughput}:
\begin{equation}
w_i = \frac{{1}/{R_i^\alpha}}{\sum \limits_j {1}/{R_j^\alpha}}, \alpha \ge 0,
\end{equation}
where \(R_i\) is the average rate of user \(i\).
The period of running the power control algorithm can be arbitrarily large, given that channel conditions do not change significantly over time. Thus TDMA can be implemented without hardware modifications required for realtime computations.

\section{Numerical Results}
\label{sec:results}

We use the NS-3~\cite{ns-3} network simulator to evaluate the developed solutions. In the considered scenario, 7~access points (APs) are arranged in a hexagonal grid with inter-site distance ISD (one AP is located in the center and six others are the edge ones), see Fig.~\ref{fig:topology}.
The APs are placed at the height of 6\,m and transmit data to corresponding STAs which are placed at the height of 1\,m.
We consider that the pathloss \cite{simulation_scenarios} is calculated as
\begin{equation*}
    \label{eq:pathloss}
    \begin{split}
        d(r) &= 40.05 + 20 \log_{10}(f_c / 2.4) \\
        &+ 20 \log_{10} (\min(r, 10)) + \indicator{r > 10} \cdot 35 \log_{10} 0.1 r,
    \end{split}
\end{equation*}
where $f_c = 5.21$\,GHz, $\indicator{r > 10}$ is an indicator function equal to $1$ if $r > 10$ and $0$ otherwise.

Data rates are calculated according to the 802.11ac model implemented in NS-3, and Minstrel~\cite{minstrel} rate control algorithm is used. However, for the purposes of utility function calculation we approximate the SINR--MCS correspondence with the following function (see Fig.~\ref{fig:rate}(a)):
\begin{equation*}
    f(x) = r(y=10 \log_{10}(x)) = \begin{cases}
		\frac{L}{1 + e^{-k(y - y_0)}},       & y \ge y_0, \\
		\frac{L}{2} + \frac{k}{4} (y - y_0), & y < y_0,
    \end{cases}
\end{equation*}
where \(y\) is the SINR in dB, \(L = 51.8\,\text{Mbit/s}\), \(y_0 = 10\,\text{dB}\), \(k = 0.17\,\text{dB}\). The part of the function to the left of \(y_0\) is replaced with a linear function to avoid overestimating capacity. Note that for the accurate optimization, we consider real data rates instead of PHY nominal data rates, i.e. we take into account interframe spaces, acknowledgments, etc.
In \eqref{eq:utilit}, we select \(\alpha = 1\), i.e. we maximize the geometric mean data rate.

Fig.~\ref{fig:throughput}(b,c) shows the geometric mean and the arithmetic mean data rate for three cases: legacy Wi-Fi behavior without power control, pure power control (i.e. power control without actions described in Section \ref{sec:dynamic}) and joint power control and scheduling.
As shown in this figure, the geometric mean and the arithmetic mean data rate turn out to be rather close to each other, which confirms that the proposed solution is fair.
When inter-site distance is very high, links do not interfere, so the maximum throughput is achieved on all links. As inter-site distance becomes lower, advantage of our algorithm becomes more apparent, doubling the geometric mean throughput. Note that when the inter-site distance is close to zero, power control cannot give all STAs at least minimal throughput simultaneously, and time division is required. In this case, both joint power control with scheduling and dynamic scheduling enable a transmission from the only one AP at a time with the maximum power.

The most interesting case is when inter-site distance is close to 20\,m. In this case, solutions involving scheduling enable simultaneous transmissions on various subsets of non-neighbor APs, thus achieving higher throughput and better fairness.

\section{Conclusion}
\label{sec:conclusion}
In the paper, we consider the problem of interference mitigation in dense Wi-Fi networks. We have developed a power control algorithm which takes into account the CST restrictions and significantly increases network throughput for middle and large inter-site distances while providing fair channel time allocation. Furthermore, by combining it with time division, we have designed a solution that provides the best performance for the whole range of inter-site distances.
In future work we plan to evaluate the proposed algorithm in more complex scenarios with frame aggregation and block acknowledgment.

\bibliographystyle{IEEEtran}
\bibliography{article}

\end{document}